\documentclass[aps,twocolumn]{revtex4}
\begin{document}

\def\StartPage{1}
\def\EndPage{6}
\pagestyle{myheadings}

\title{ Maxwell's Demon, Rectifiers, and
the Second Law: Computer Simulation of Smoluchowski's Trapdoor}

\author{P. A. Skordos}
\affiliation{Santa Fe Institute, Santa Fe, NM}
\affiliation{Massachusetts Institute of Technology, 545 Technology Square, Cambridge, MA 02139}

\author{W. H. Zurek}
\affiliation{Theoretical Division, MS B213, Los Alamos National Laboratory, Los Alamos, NM 87545}
\date{December, 1991}

\begin{abstract}
We have simulated numerically an automated version of Maxwell's demon inspired by
Smoluchowski's ideas of 1912.  Two gas chambers of equal volume are
connected via an opening that is covered by a trapdoor.  The trapdoor
can open to the left but not to the right, and is intended to rectify
naturally occurring fluctuations in density between the two chambers.
Our results confirm that though the trapdoor behaves as a rectifier
when large density differences are imposed by external means, it can
not extract useful work from the thermal motion of the molecules when
left on its own.
\end{abstract}

\maketitle
\markboth{\sl }{\sl }

\section{Introduction}

The second law of thermodynamics has been a subject of debate ever
since it was formulated. It says that the entropy of a closed
system can only increase with time, and thus natural
phenomena are irreversible.  In other words a system left
on its own can only evolve in one direction, towards equilibrium.
This is in contrast to time-reversible dynamics and raises the
question of how reversible dynamics can lead to macroscopic
irreversibility.  An answer can be furnished using probabilistic
arguments in statistical mechanics, but the arguments are difficult to
translate into a rigorous proof without postulating a new axiom about
nature, the {\em stosszahlansatz}, also called the assumption of
molecular chaos \cite{davies-pc,ehrenfest}, which is at odds with
dynamic reversibility.  As a result, the origin of the second law of
thermodynamics remains elusive and provides a source of interesting
discussions on the foundations of physics.

A popular way of challenging the second law is the idea of ``perpetual
machine of the second kind'', which is supposed to extract useful work 
in a closed cycle from the perpetual thermal motion of gas molecules.  This is
prohibited by the second law because if it were possible to convert
thermal energy into useful work in a closed, equilibrated system, 
then the entropy of an isolated system could be made to decrease.
However, it remains puzzling why a microscopic device can not be constructed 
to take advantage of spontaneous variations in density between microdomains of
gas, to bring a system from a state of maximum disorder (equilibrium)
into an ordered state, and eventually to convert thermal energy into
useful work.

The history of microengines that convert thermal energy into useful
work began when J.C. Maxwell proposed such a microengine at the end of
his book {\em Theory of Heat}~\cite{ehrenberg_sci_am}, which he named
a ``demon''. Since then the term ``demon'' has become a standard.
Maxwell's demon works by opening and closing a tiny door between two
gas chambers, based on the information that the demon has about
individual molecules.  The method used to obtain information is not
specified.  The demon allows only fast molecules to pass from left to
right, and only slow ones to pass from right to left.  This leads to a
temperature difference between the two gas chambers, which can be used
to convert thermal energy into useful work.  It is currently believed,
however, that Maxwell's demon can not violate the second law
\cite{bennett-sci-am,szilard,brillouin,demonology} because the
information needed to operate the demon's door does not come without a
price.  As Bennett explains in \cite{bennett-sci-am} following an idea
that goes back to \cite{szilard} and \cite{landauer-ibm}, the demon
must dissipate energy into a heat bath in order to erase the
information that it obtains by examining molecules.  The heat bath may
be the molecules that the demon examines or it may be another system
that is colder than the demon and the gas molecules. The energy lost
in the heat bath is always greater than or equal to the energy that
can be extracted after the demon has finished its operations.  Thus,
the second law of thermodynamics is not
violated even when the demon is ``intelligent" --- that is, capable
of universal computing ability --- as it was pointed out by one of
us~\cite{zurek,caves}.

There is an alternative approach, however, to designing microengines
that convert thermal energy into useful work; and that is to focus on
purely mechanical devices, and to avoid the issues of measurement and
information that haunt the original Maxwell's demon.  Purely
mechanical, automated Maxwell's demons are completely described within
a theoretical framework such as newtonian physics. In particular, there is no
measurement mechanism outside of such mechanical model.  An example is the
trapdoor mechanism discussed by Smoluchowski in
\cite{smoluchowski1,smoluchowski2}. A well-known and similar in the spirit to
Smoluchowski's trapdoor is
the ratchet and pawl mechanism discussed by Feynman in \cite{feynman}.

Our paper describes the computer simulation of a trapdoor mechanism
inspired by Smoluchowski's ideas.  Our results confirm Smoluchowski's
insight that though the trapdoor acts as a rectifier when large
density differences are imposed by external means, it can not extract
useful work from the thermal motion of the molecules when left on its
own.  The next section describes our trapdoor mechanism and the
simulation program. Following that, we discuss how the trapdoor
succeeds at rectifying large density differences that are imposed by
external means.  Then, we show how our trapdoor system fails to work
when left to operate on its own. Finally, we discuss how the trapdoor
can be modified to work successfully as a pump and create density differences
in the system initialy in equilibrium. This requires, however, dissipation
of its random motions which can be accomplished only by ``opening" the
system, that is, for example, by keeping the trapdoor at a lower temperature 
than the molecules. Such a pump, of course, is no more a threat to the 
second law than is a refrigerator.

\section{Description of the Model}

The system that has been simulated is shown in Figure
\ref{fig:demon1}. It consists of two gas chambers of equal area,
connected via an opening that is covered by a trapdoor.  The
simulation is two-dimensional and the gas molecules are billiard balls
moving on the plane and colliding with each other.  All the collisions
conserve energy and momentum, except for particle-wall collisions that
reflect a particle's momentum like light rays reflect off mirrors. The
collision forces are derived from infinite hard core potentials, and
angular momentum (spin of the billiard balls) is not included in the
model.  All collision forces act radially through the center of the
colliding balls.

The trapdoor is constrained to move between two {\it door stops}, one
of which is located at the middle wall, and the other is located
inside the left chamber.  The location of the door stops allows the
trapdoor to open to the left but not to the right.  This endows the
trapdoor with rectifying properties, as we shall see in the next
section.  The trapdoor is a line segment of zero width, impenetrable
by the molecules, which moves horizontally at constant speed, and
reverses its direction when it comes in contact with the door stops.
During its motion, the trapdoor slides along ideal rails and thus
remains vertical at all times. Accordingly, collisions between the
trapdoor and the molecules conserve energy and momentum except for
collisions at the edges of the trapdoor which do not conserve the
y-component of momentum (that is, the trapdoor does not move at all in that
direction  --- it can be thought of as moving on rails which are attached to 
the infinitely massive box containing the gas). The collision equations 
are discussed in detail below.

To evolve the system of molecules and the trapdoor we use the
following algorithm: Given the positions and velocities at time $t_0$,
we find all the collisions that are about to occur.  We select the
shortest collision time $\Delta t$, and move all the particles and the
door freely during $\Delta t$. At time $t_0 + \Delta t$ we perform the
collision that occurs at this time, and then repeat the cycle looking
for the next shortest collision time.  The algorithm works because
collisions in our system are instantaneous.  The types of collisions
that can occur are of four types: particle against particle, particle
against wall, particle against door, and door against door stop.

The evolution algorithm can be implemented efficiently on a computer
if we are careful to avoid unnecessary computations.  For example, we
do not need to examine all pairs of particles at every time step.  If
we see that two particles are far from each other, then we need not
examine them again until a number of time steps have elapsed.  Only
then these two particles will have another chance of being near each other
and being able to collide.  Also, if we compute the collision time for
a pair of particles that are near each other, and another pair of
particles collides before them, we need not discard the first
collision time.  We simply decrement the first collision time by the
time interval by which the whole system is evolved.  A word of caution, 
however, is necessary.
The process of decrementing collision times should not be repeated
more than a few times because the roundoff error in subtracting small
intervals of time becomes significant very quickly.  Also, a collision
involving some particle must invalidate all pre-computed collision
times involving that particle.

Two kinds of elementary formulas are used in the evolution algorithm:
{\it collision equations} give the new velocities in
terms of the old velocities, and {\it timing equations} give the
time interval until an upcoming collision.  The timing equations are
the simpler of the two. They are derived from geometrical
constraints and the fact that the particles and the door move at
constant velocity between collisions.  For example, to compute the
collision time between two particles, we ``draw a straight line" from the
current position of the particles to the point where the particles are
tangential to each other. The geometric constraint of tangency allows
two possibilities, and we have to choose the one that occurs first and
is the physical one.  Algebraically we have to solve a quadratic
equation, and to pick the smallest positive solution.  The timing
equations for the other types of collisions in our system are derived
in a similar manner.

The collision equations are a little more complicated than the timing
equations.  As usual the collision equations are derived from
conservation of kinetic energy, conservation of linear momentum, and
the condition that forces act radially.  The last condition means that
the force vector must pass through the center of the particle disk
that is colliding, and hence momentum is exchanged along this
direction. For nearly all collisions the radial force condition is satisfied
automatically in setting up the geometry of the problem. However,
there is one type of collision in our system that requires explicit
use of the radial condition.  This occurs when a particle disk
collides with the edges of the moving trapdoor.  Since it is not
discussed in most textbooks, we review briefly the equations.

The radial action condition requires that the change in y-momentum
divided by the change in x-momentum equals the tangent of the angle
$\theta$ formed by the center of the colliding disk, the point of
contact, and the x-axis.  The point of contact is the edge of the
moving trapdoor.  If $v_x, v_y$ are the old velocities and $v_x',
v_y'$ are the new velocities of the colliding disk, we have the
equation
\begin{equation}
(v_x - v_x') = \frac{\cos\theta}{\sin\theta} (v_y-v_y') \;\;\;.
\label{eqn:radial-action}
\end{equation}
To find the velocities following a collision in terms of the
velocities before the collision, we use
equation~\ref{eqn:radial-action} together with kinetic energy and
x-momentum conservation.  The y-momentum is not conserved because the
trapdoor moves on ideal rails, and its y-velocity is always zero.
After some algebra we get the following equations for the new
velocities,
\begin{equation}
v_x' = \frac{-2cs v_y + 2 c^2 V_x + (s^2-\delta c^2) v_x}{(\gamma
c^2 + s^2)} 
\end{equation}
\begin{equation}
v_y' = \frac{-2cs v_x + 2 cs V_x + (\gamma c^2- s^2) v_y}{(\gamma
c^2 + s^2)} 
\end{equation}
\begin{equation}
V_x' = V_x + \frac{m}{M} (v_x-v_x')
\end{equation}
where
\begin{displaymath}
\begin{array}{lll}
c &=& \cos \theta	\\
s &=& \sin \theta	\\
\gamma &=& (1 + m/M)	\\
\delta &=& (1 - m/M)	\\
\end{array}
\end{displaymath}
and where $M, V_x$ are the mass and x-velocity of the trapdoor; $\;
m,v_x,v_y$ are the mass and velocities of the particle; and $\theta$
is the angle formed by the center of the particle, the colliding edge
of the trapdoor, and the x-axis.  The collision equations for all
other types of collisions in our system can be found in standard
textbooks~\cite{loeb}.

The numbers we used in our simulations were chosen to correspond to a
standard gas like nitrogen.  We experimented with different values for
the size of the gas chambers, molecular speeds, and other quantities,
and the qualitative behavior of the gas was the same for all choices.
We looked at systems containing a number of molecules ranging from 20
to 500.  We chose the radius of the molecules to be $3\times
10^{-8}cm$, mass $4.7\times 10^{-23}gm$, and velocities of the order
$10^4 cm/sec$.  We chose the size of the gas chambers to give a mean
free path between collisions of the order of the size of the chambers.
Specifically in the case of 500 molecules, the width of the each
chamber was $13.5\times 10^{-6}cm$ and the height was $18\times
10^{-6}cm$. The mean free path at equilibrium in each chamber is
estimated by the ratio \cite{footnote-mean-free-path},
\begin{displaymath}
\lambda = \frac{\mbox{Area}}{n\; \times 2\; R}
\end{displaymath}
which gives $\lambda = (13.5 \times 18 \times 10^{-12})/(250 \times 2
\times 3 \times 10^{-8}) = 16.2 \times 10^{-6} cm$.  

We experimented with different masses for the trapdoor, and in the
results reported below the mass of the trapdoor is of the order of
three to four times the mass of one particle, unless otherwise
indicated. Using a trapdoor mass which is comparable to the molecular
mass leads to an average speed for the trapdoor that is comparable to
that of the molecules, and facilitates numerical simulation.  A very
light trapdoor moves too fast and increases the numerical roundoff
error; while an excessively heavy trapdoor (orders of magnitude
heavier than a molecule) can delay the approach to equilibrium and
requires longer time averaging. It should be emphasized that our 
qualitative conclusions are expected to be independent of the mass of the 
trapdoor, and we certainly do not have any indications in our numerical results
to question this expectation. Thus, the choice of the mass of the door was
dictated primarily by the above considerations of numerical convenience. 
Our simulations took typically several days using standard Unix workstations,
and the relative error in the total energy of the system was kept less 
than $10^{10}$ using double precision arithmetic. The running times 
were dictated primarily by the desire to gather good statistics.
By contrast, the rectifying behavior described in the next section can
be seen on a much less imposing timescale of a few minutes, depending of 
course on the efficiency of the algorithm and on the number of particles
used. 

\section{The Trapdoor as a Rectifier}
\label{sec:rectifier}

This section discusses the behavior of the trapdoor when large density
differences between the two chambers are imposed by external means.
It is found that under these circumstances the trapdoor acts as a
diode, and prolongs the duration of states of higher density in the
left chamber.  There are a number of ways to exhibit this rectifying
behavior. We shall describe here three of them.

The first way is to measure the equilibration time or transient
response to an initial density difference, for example when all
molecules start in the left chamber.  To be precise we place all
molecules along the outermost wall of one chamber with the trapdoor
set motionless in the closed position, and measure the density in each
chamber until the populations in the two chambers equalize.  Figure
\ref{fig:rt1} shows the absolute value of the difference in the number
of molecules between the two chambers plotted against time.  The
difference in the number of molecules is normalized by the total
number of molecules, which is $500$ in this experiment.  Two curves
are shown, one for the case when all molecules start in the left
chamber, and one for the case when all molecules start in the right
chamber. We see that in the latter case the populations equalize
``immediately". In other words the density difference vanishes much more
quickly when the molecules start in the right chamber than when the
molecules start in the left chamber.

The second way of observing the rectifying behavior of the trapdoor is
shown in Figure \ref{fig:rt1b}.  The data comes from the same kind of
equilibration experiment as Figure \ref{fig:rt1}, where all the
particles are positioned initially along the outermost wall of one
chamber. Now, we look at the time interval it takes $25$ molecules to
pass from one chamber to the other as a function of the density
difference. If $T$ is this time interval, then the ratio $25/T$
measures the current of particles --- the rate --- at which they pass through 
the middle wall opening in response to the density difference during that time.
Figure \ref{fig:rt1b} plots the particle current against the density
difference between the two chambers for a system of $500$ particles.
The resulting curve resembles qualitatively the voltage-current
characteristic of an electrical diode, and indicates that the trapdoor
acts as a rectifier when large density differences are imposed by
external means \cite{ehrenberg_sci_am,brillouin-2}.

The precise quantities plotted in Figure \ref{fig:rt1b} are the rates ---
inverse time intervals $1/(T_{i+1}-T_i)$ --- as a function of the average
number of particles difference between the two chambers $(\Delta
N_{i+1} + \Delta N_{i})/2\;$ which exists during the interval
$(T_{i+1}-T_{i})$.  $\Delta N_i$ is the number of particles difference
at the starting time $T_i$, and $\Delta N_{i+1}$ is the number of
particles difference at the finishing time $T_{i+1}$ when $25$
molecules have moved from the source chamber (high density) to the
sink chamber (low density).  The y-axis is in units of $25\times
10^{9} \mbox{particles}/sec$.  The x-axis is in units of the
normalized difference in the number of particles, so that an interval
of size $0.1$ corresponds to $25$ particles moving from one chamber to
the other $(0.1 \times 500 \times \frac{1}{2} = 25)$.  The intervals
$(-1,-0.9)$ and $(0.9,1.0)$ are not included in the plot because the
times immediately after the release of the system from our initial
conditions do not correspond to smooth flow.

The third and last method of exhibiting the rectifying behavior of the
trapdoor focuses on steady state behavior.  In contrast to the
equilibration experiments above, this method measures the time averaged
flow of molecules through the middle wall opening when a large density
difference is maintained artificially.  In particular, we continualy ``reverse 
bias" the system by removing the molecules that hit the rightmost wall of
the right chamber and reinserting them in the left side of the left
chamber. This results in a density difference that pushes the trapdoor
towards the closed position.  In an opposite experiment we forward
bias the system by reinserting molecules from the leftmost wall into
the right chamber.

\begin{table}[htp]
\begin{center}
\begin{tabular}{|c|r|r|}
\hline
\mbox{ }	&\mbox{N = 500}	&\mbox{N = 100}	\\
\hline
\mbox{Reverse Bias}	&$4.26\times 10^9$	&$1.05\times 10^9$	\\
\mbox{Forward Bias}	&$-9.31\times 10^{10}$	&$-1.81\times 10^{10}$	\\
\mbox{ratio}	&$1:22$		&$1:17$			\\
\hline
\end{tabular}
\end{center}
\caption{ The flow of molecules through the middle wall opening in
forward and reverse bias conditions, for systems of $100$ and $500$
molecules. The molecules crossing from left to right are counted
positive, and those crossing from right to left are counted negative.
}
\label{rt2}
\end{table}

Table~\ref{rt2} lists the flow of molecules (number of particles per
second) passing through the middle wall opening under reverse and
forward bias conditions.  Molecules passing left to right are counted
positive and molecules passing right to left are counted negative.  We
list the results for two different systems, a system of $500$
particles and a system of $100$ particles.  The time intervals used to
time averaged are about $10^{-5} sec$ for the $100$ particle system and
$10^{-6} sec$ for the $500$ particle system, which are both large
enough to guarantee convergence; that is the average values will not
change over longer time intervals.  We have checked this by plotting
the time averages against time, and seeing that the curves approach a
horizontal slope and a constant value.  The values in table~\ref{rt2}
show that the flow allowed by the trapdoor in the forward bias
condition is $22$ times as large as the flow allowed in the reverse
bias condition for $500$ particles, and $17$ times as large in the
case of $100$ particles.  In other words, the trapdoor acts as a
rectifier.

It is worth pointing out that the rectifying behavior of the trapdoor
depends greatly on the geometry of the system.  Experimentally we have
found that our trapdoor becomes a better rectifier the longer the
trapdoor is, and the more molecules there are near the trapdoor.  When
many collisions take place exclusively on one side of the trapdoor
during the time interval it takes the trapdoor to move significantly,
the trapdoor is pushed and kept near one door stop.  The probability
of moving significantly away from that door stop is very small.  For
example, if many collisions take place exclusively on the left side of
the trapdoor, the trapdoor will be kept near the middle wall opening,
bouncing between the middle wall door stop and the large number of
particles on its left side.  The trapdoor performance can be improved
further by placing one door stop slightly inside the right chamber.
This centers the jittering of the door exactly on the middle wall and
decreases the chance of a molecule leaking from the high density left
chamber into the low density right chamber. For similar reasons we
expect that making the trapdoor have finite width, that is using a two
dimensional trapdoor in the shape of a rectangle will result in even
better rectifying behavior for large density differences.

\section{Verification of the Second Law}

Having seen that the trapdoor acts as a rectifier under external bias,
we now consider what happens when the trapdoor and molecules are left
to evolve on their own in an isolated container.  The asymmetry of the
trapdoor's location, opening to the left but not to the right, intends
to hinder the passage of molecules from left to right while providing
an easy access from right to left.  In this way, the trapdoor attempts
to exploit the naturally occurring fluctuations in density between the
two chambers and to make states of high density in the left chamber
last longer than corresponding states of high density in the right
chamber.  Ultimately, the trapdoor attempts to keep a higher average
number of molecules in the left chamber than in the right chamber.

However, our simulations show that the trapdoor does not succeed.
When the trapdoor and molecules are left to evolve on their own, the
time average number of molecules in the left chamber is actually
smaller than the time averaged number of molecules in the right
chamber.  Moreover, this imbalance is not a true density difference
and does not violate the second law.  The reason for the unequal
number of molecules is that the presence of the trapdoor in the left
chamber occupies space, which makes the available area in the left
chamber slightly smaller than the available area in the right chamber.

The effect of the excluded area by the trapdoor has been measured by
performing an experiment of $20$ particles where each chamber measured
$13.5\times 10^{-7}cm$ horizontally and $18\times 10^{-7}cm$
vertically.  The particle radius was $R = 6\times 10^{-8}cm$ giving a
mean free path in the order of $20\times 10^{-7}cm$.  The length of
the trapdoor (vertical direction) was $10\times 10^{-7}cm$.  Given
these numbers we can estimate the average number of particles in the
left chamber by assuming uniform density (equilibrium) in a time
average sense.  If $N_L$ is the time averaged number of particles in
the left chamber and $A_L$ the available area in the left chamber, we
have
\begin{displaymath}
\frac{N_L}{A_L} = \frac{N_R}{A_R} = \frac{1-N_L}{A_R}
\end{displaymath}
which gives $N_L = A_L/(A_L + A_R)$.  To estimate the available area
in each chamber, $A_L$ and $A_R$, we consider the area in which the
{\it centers} of the particles can travel. Thus 
$A_R = \left[ (13.5 \times 18) - (13.5 + 13.5 + 18)\times 0.06
\right] \times 10^{-14}cm^2$ for the right chamber, and for the left chamber we subtract from the
above $A_R$ the area excluded by the trapdoor $\left[ (10 \times 0.06)
+ \pi (0.06)^2 \right] \times 10^{-14}cm^2$.  Putting these together
we find $N_L = 0.484$.  In simulating this experiment we found the
time averaged number of particles in the left chamber to be $0.486$ in
excellent agreement with the theoretical estimate
\cite{footnote-excluded-volume}. 

There is also an alternative way of checking that the observed 
particle number difference between the two chambers is not a true density
difference, and it can not lead to a violation of the second law.  The
idea is to open a second hole in the middle wall, in addition to the
opening covered by the trapdoor. If the trapdoor could lead to a true
density difference, pumping molecules from one chamber to the other
through the opening covered by the trapdoor, then the second free
opening should exhibit the {\it return flux} of molecules and lead to
perpetual flow between the two chambers.  Our simulations of this
experiment did not show any flow.

Our simulations show that the operation of the trapdoor is consistent
with the second law of thermodynamics in the sense that the particles
are distributed uniformly in the available area on the average, and
the entropy of the system is maximized.  Our simulations have also
shown that the time averaged temperature is the same in each chamber
and equal to the temperature of the trapdoor (average kinetic energy
divided by the number of degrees of freedom, two for the particles,
and one for the trapdoor). Finally, we have confirmed that the time
averaged velocities in each chamber are distributed as Gaussian
distributions in $v_x$ and $v_y$. The two Gaussian distributions are
identical to each other and identical between the two chambers, which
is consistent with the Maxwell-Boltzmann distribution law and
equipartition of energy.

An intuitive explanation of why our trapdoor system fails to work when
operating in an isolated container of gas molecules is that the
trapdoor gets thermalized --- its temperature equals the temperature
of the particles --- and the trapdoor's thermal motion prevents the
rectifying behavior \cite{smoluchowski1,smoluchowski2,feynman}.  To contrast, a
macroscopic trapdoor works successfully as a rectifier because it can
get rid of excess energy through dissipation.  Following this analogy
a little further, we may expect that our trapdoor would work
successfully if a reservoir of lower temperature than the particles
were used to dissipate its motion.  The trapdoor would then act as a pump,
letting the molecules through from one chamber to the other
more easily in one direction than the other, on the average.
We have
tested this idea in simulations, and we report the results in the next
section.

\section{The Trapdoor as a Pump}

To convert our trapdoor system into an effective pump, we modify the
evolution algorithm to cool the trapdoor by removing energy in small
increments.  In particular, we scale the trapdoor's velocity by $0.5$
every $\Delta t$ time interval with $\Delta t$ sufficiently small.
The lost energy is reinserted in equal amounts to all the particles by
scaling their velocities, conserving the total energy of the system.
Moreover, the cooling of the trapdoor is performed only when the
trapdoor is near the closed position, which makes the trapdoor tend to
remain closed. The goal of the cooled trapdoor is to pump molecules
from the right chamber into the left chamber.

Our experiments show that this design works successfully. Further,
they show that the mass of the trapdoor in relation to the mass of
each particle is crucial for efficient operation.  If the trapdoor
mass is much smaller than the mass of one particle, then the action of
a single particle coming from the right chamber is enough to open the
trapdoor and let the particle through, even though some energy is lost
by interacting with the trapdoor.  Easy access from the right chamber
is desirable.  On the other hand, if the trapdoor mass is much larger
than the mass of one particle, many particles must collide with the
trapdoor in a short period of time in order to open the trapdoor.
Clearly, this situation does not occur very frequently, and so a heavy
trapdoor does not work very well.  Our simulations show that a very
light trapdoor with dissipation can act effectively as a one-way door,
opening to particles from the right, and remaining closed to particles
from the left.  A heavier trapdoor with dissipation works also, but
not as well.

In Figure \ref{fig:heat-engine} we report results for a trapdoor
system with dissipation, where the mass of the trapdoor is $4.7\times
10^{-24}gm$, or one tenth of the mass of one particle.  The time
interval which controls the rate of energy dissipation is $2.5\times
10^{-13}sec$, while the mean free path and mean collision time in the
left chamber are of the order of $20\times 10^{-7}cm$ and $5\times
10^{-11} sec$.  The length of the trapdoor is $6\times 10^{-7}cm$ and
each chamber measures $13.5\times 10^{-7}cm$ horizontally and
$18\times 10^{-7}cm$ vertically.  In this experiment we have also
included a second hole in the middle wall, of size $1\times
10^{-7}cm$, in addition to the hole covered by the trapdoor.  The
purpose of the additional hole is to verify that the trapdoor indeed
acts as a pump of molecules from right to left, by exhibiting the {\it
return flux} of molecules left to right.  Graph (a) of Figure
\ref{fig:heat-engine} shows how the normalized number of molecules in
the left chamber builds up as soon as the system is released from a
state of equal number of molecules in each chamber.  Similar
simulations that were run much longer than Figure
\ref{fig:heat-engine} show that the time averaged of the number of
molecules in the left chamber stabilizes around $0.76\;\;$ (over
$10^{-5} sec$).  Graph (b) of Figure \ref{fig:heat-engine} shows the
accumulated flux of particles through the trapdoor covered opening,
and the accumulated flux of particles through the second opening that
allows free passage.  The slope of the accumulated flux (measured by
averaging over $10^{-5} sec$) is approximately $2\times
10^{9}\mbox{particles}/sec$.  The time averaged temperature of the
trapdoor is $11$ degrees Kelvin, compared to $270$ degrees Kelvin for
the particles. These results show that the trapdoor can operate
successfully as a rectifier when a reservoir of lower temperature is
available. However, as discussed previously it can not operate
successfully when run at the same temperature as the gas molecules, in
accordance with the second law of thermodynamics.

\newpage


\newpage 


\begin{figure}[htbp]
\caption{ The automated Maxwell's demon we simulated numerically
was inspired by Smoluchowski's trapdoor. The dashed lines show the
region where the trapdoor can move. }
\label{fig:demon1}
\end{figure}

\begin{figure}[htbp]
\caption{ The normalized absolute value of the difference between the
number of particles in the two chambers is plotted against time, as
the system approaches equilibrium. Two curves are shown, one for the
case when the particles start in the left chamber, and one for the
case when the particles start in the right chamber. }
\label{fig:rt1}
\end{figure}

\begin{figure}[htbp]
\caption{ The flux of particles from one chamber to the other is
plotted against the density difference that initiates the flux.  $N1$
is the number of particles in the left chamber, $N2$ the right
chamber, and $N$ the total number of particles.  The y-axis is in
units of $25\times 10^{9} \mbox{particles}/sec$. }
\label{fig:rt1b}
\end{figure}

\begin{figure}[htbp]
\caption{ A trapdoor with a cooling mechanism acts as a pump.
The graph on the top (a) shows how the fraction of the total number of molecules
in the left chamber builds up after the system is released from an
initial state with an equal number of molecules in each chamber. The
graph on the bottom (b) shows the accumulated flux of molecules
through the trapdoor opening (negative slope) and the accumulated flux
of molecules through an additional opening that allows free passage
(positive slope). }
\label{fig:heat-engine}
\end{figure}

\end{document}